\documentstyle[12pt]{article}
\def\beq{\begin{equation}}
\def\eeq{\end{equation}}
\def\bea{\begin{eqnarray}}
\def\eea{\end{eqnarray}}
\def\bq{\begin{quote}}
\def\eq{\end{quote}}

\def\nnb{\nonumber}
\def\ga{\left(}
\def\dr{\right)}
\def\aga{\left\{}
\def\adr{\right\}}
\def\als{\alpha_s}

\def\rar{\rightarrow}
\def\nnb{\nonumber}

\def\nin{\noindent}

\newcommand{\AmS}{{\protect\the\textfont2
  A\kern-.1667em\lower.5ex\hbox{M}\kern-.125emS}}
\begin{document}
\topmargin -1.5cm
\oddsidemargin +0.2cm
\evensidemargin -1.0cm
\begin{flushright}
{CERN-TH.7506/94}\\
PM 94/43
\end{flushright}
\vspace*{5mm}
\begin{center}
\section*{$\alpha_s$ \bf
 FROM TAU DECAYS$^{*)}$}
\vspace*{0.5cm}
{\bf S. Narison} \\
\vspace{0.3cm}
Theoretical Physics Division, CERN\\
CH - 1211 Geneva 23\\
and\\
Laboratoire de Physique Math\'ematique\\
Universit\'e de Montpellier II\\
Place Eug\`ene Bataillon\\
34095 - Montpellier Cedex 05\\
\vspace*{1.5cm}
{\bf Abstract} \\ \end{center}
\vspace*{2mm}
\noindent
We review the present status in the determination of the $accurate$
value of $\alpha_s$ from $\tau$-decays, where we discuss in
detail  the different sources of theoretical errors.

\vspace*{4.0cm}
\noindent
\rule[.1in]{16.0cm}{.002in}

\noindent
$^{*)}$ Talk given at the third Workshop on Tau Lepton Physics, Montreux,
Switzerland, September 1994. To appear in Nucl. Phys. B, proceedings
supplement.
\vspace*{0.5cm}
\noindent


\begin{flushleft}
CERN-TH.7506/94 \\
PM 94/43\\
December 1994
\end{flushleft}
\thispagestyle{empty}
\vfill\eject

\setcounter{page}{1}
\section{INTRODUCTION} \par
\nin
Since the original and previous theoretical works in
\cite{BRA}-\cite{BRAT} and the
ALEPH measurement\cite{ALEPH}, there has been an amount of progress both
theoretically
and experimentally in the determination of $\als$ from tau decays. The
theoretical
progress resides mainly in a better understanding of the QCD perturbative
 and
nonperturbative contributions to the inclusive hadronic width, as
discussed in recent reviews\cite{PICHM}-\cite{SN94} while the
experimental
one can be essentially due to the improved measurement of the leptonic
width (see
e.g. \cite{DUFLOT}) and of the $\tau$-lifetime.
\nin
In this (written) version of my talk, I will only limit to the discussion
 of
the different sources of
theoretical errors in the determination of $\als$, in order to avoid a
repetition of
the different discussions already done in previous works. In this way,
the present paper
complements the previous review in \cite{SN94}.
\nin
It is now well-known that the inclusive branching ratio:
\beq
 R_\tau   \equiv  {\Gamma \left(\tau  \longrightarrow  \nu
_\tau + \mbox{hadrons} \right) \over \Gamma \left(\tau  \longrightarrow
 \nu_ \tau
e\bar  \nu_ e \right)},
\eeq
can be reliably measured through the leptonic branching ratio $B_l$:
\beq
 R^B_\tau   \equiv  {1 - B_e- B_\mu \over B_e}    .
\eeq
or/and the $\tau$-lifetime:
\beq
 R^\Gamma_ \tau    \equiv  {\Gamma_ \tau - \displaystyle \sum^{
}_{ e,\mu} \Gamma_{ \tau  \longrightarrow \nu_ \tau } \over \Gamma_{ \tau
\longrightarrow \nu_ \tau} }
\eeq
The PDG94 compilation gives the value \cite{PDG}:
\beq
R^\Gamma_\tau=3.55 \pm 0.06~~~~~~~~~~
R^B_\tau=3.56 \pm 0.04,
\eeq
leading to the average:
\beq
R_\tau=3.56 \pm 0.03.
\eeq
However, there is a continuous effort in improving the measurements
of the leptonic width in the four LEP experiments as presented in
this workshop. The most precise
recent preliminary data come from
the ALEPH group and lead to\cite{DUFLOT}:
\beq
R^B_\tau=3.645 \pm 0.024,
\eeq
which are in agreement with the other LEP data \cite{L3}, but
 are larger by about $2\sigma$ than the CLEO II result \cite{CLEO},
using PDG values of the leptonic width:
\beq
R^B_\tau=3.552 \pm 0.035.
\eeq
These data are larger than the na\"{\i}ve parton model prediction
$R_\tau= N_c$ by about 20$\%$. This discrepancy is expected to be
resolved by the
electroweak and QCD corrections.
\nin
Using the analyticity of the two-point correlator built from the vector
and
axial-vector currents, which gouverns the semileptonic process, and using
 the Cauchy
theorem which tranforms the integral over the real axis into
the one around the circle of a radius $M_\tau$ in the complex $s$-plane,
the QCD expression of the inclusive width can be written as \cite{BNP}:
\bea
 R_\tau   &\equiv&  3 \left( \left\vert V_{ud} \right\vert^ 2+
\left\vert V_{us} \right\vert^ 2 \right)\ S_{EW}\nnb \\
&& \left\{ 1 + \delta_{ EW}+
\delta^{ (0)}+ \sum^{ }_{ D=2,4,...}\delta^{ (D)} \right\}  ,
\eea
\par
\smallskip
\noindent where $ \left\vert V_{ud} \right\vert
\simeq  0.9744 \pm  0.0010$, and
$
 \left\vert V_{us} \right\vert  \simeq  0.2205 \pm  0.0018 $ are the CKM
mixing angles \cite{PDG}; $S_{EW}=1.0194$ \cite{MARC} and $\delta_{EW}=
0.0010$ \cite{LI}
are the electroweak corrections coming respectively from the summation of
 the
leading-log and from the constant term; $\delta^{(0)}$ is the
perturbative contribution.
Within the standard SVZ-expansion \cite{SVZ},
 $\delta^{(2)}$  is the quark mass corrections, while
 $\delta^{(D\geq 4)}$ is the nonperturbative contributions simulated by
 the operators of dimension $D$.
\section{THE SIZE OF DIFFERENT CORRECTIONS}

\begin{table*}[hbt]
\setlength{\tabcolsep}{1.5pc}
\newlength{\digitwidth} \settowidth{\digitwidth}{\rm 0}
\catcode`?=\active \def?{\kern\digitwidth}
\caption{ QCD predictions for the different components
of the $\tau$ hadronic width:}
\begin{tabular}{c|c c c  c}
\hline
&&&& \\
$\alpha_s(M^2_{\tau})$&$R_{\tau,V}$
& $R_{\tau,A}$ & $R_{\tau,S}$ & $R_\tau$ \\
&&&& \\
\hline
&&&& \\
$0.20$ & $1.62 \pm 0.02$ & $1.53 \pm 0.03$ & $0.145 \pm 0.005$ &
$ 3.29 \pm 0.01 $\\
$0.22$ & $1.64 \pm 0.02$ & $1.54 \pm 0.03$ & $0.145 \pm 0.005$ &
$ 3.33 \pm 0.02 $\\
$0.24$ & $1.66 \pm 0.02$ & $1.56 \pm 0.03$ & $0.145 \pm 0.005$ &
$ 3.37 \pm 0.02 $\\
$0.26$ & $1.68 \pm 0.02$ & $1.58 \pm 0.03$ & $0.145 \pm 0.005$ &
$ 3.41 \pm 0.02 $\\
$0.28$ & $1.70 \pm 0.02$ & $1.61 \pm 0.03$ & $0.145 \pm 0.005$ &
$ 3.45 \pm 0.02 $\\
$0.30$ & $1.72 \pm 0.02$ & $1.63 \pm 0.03$ & $0.145 \pm 0.006$ &
$ 3.50 \pm 0.02 $\\
$0.32$ & $1.75 \pm 0.02$ & $1.65 \pm 0.03$ & $0.145 \pm 0.006$ &
$ 3.54 \pm 0.03 $\\
$0.34$ & $1.77 \pm 0.02$ & $1.67 \pm 0.03$ & $0.145 \pm 0.006$ &
$ 3.58 \pm 0.03 $\\
$0.36$ & $1.79 \pm 0.02$ & $1.69 \pm 0.03$ & $0.144 \pm 0.006$ &
$ 3.63 \pm 0.03 $\\
$0.38$ & $1.81 \pm 0.03$ & $1.71 \pm 0.03$ & $0.144 \pm 0.007$ &
$ 3.67 \pm 0.04 $\\
$0.40$ & $1.83 \pm 0.03$ & $1.73 \pm 0.03$ & $0.143 \pm 0.007$ &
$ 3.71 \pm 0.04 $\\
$0.42$ & $1.85 \pm 0.03$ & $1.75 \pm 0.04$ & $0.143 \pm 0.007$ &
$ 3.75 \pm 0.04 $\\
&&&& \\
\hline
\end{tabular}
\end{table*}
\par
\smallskip
\nin
The size of these different contributions has been
discussed in detail in the original paper \cite{BNP} and in different
review talks
\cite{PICHM}-\cite{SN94},
so that we will not discuss them in any more detail here. They can be
summarized as:
\subsection{Perturbative corrections}
\nin
We use the result of \cite{KAT} for
the ${\cal D}$-function gouverning the $e^+e^- \rar \mbox{hadrons}$
total cross-section:
\beq
{\cal D}(s)= \frac{1}{4\pi^2}\sum_{n=0} K_n \left({\alpha_ s \over \pi}
\right)^n,
\eeq
where for the charged vector and axial-vector current and for three
flavours:
\beq
K_0=K_1=1~~~~K_2=1.6398~~~~K_3=6.3711,
\eeq
from which, we deduce
the perturbative corrections to order $\als^3$ \cite{BNP}:
\bea \delta^{ (0)}_{BNP} = \ga{\alpha_ s \over \pi}\dr  &+& 5.2023\
\left({\alpha_ s \over \pi} \right)^2\nnb \\
&+& 26.366\ \left({\alpha_ s \over \pi}
\right)^3
\eea
 which, for a typical value of $\als(M_\tau)$ gives a correction of about
  20$\%$, while
each term is respectively 11\%, 6\% and 3\%, indicating that at this low
$\tau$-scale the
convergence of the
calculated perturbative series is quite good, though it can be improved
\cite{LEDI} by
using another expansion parameter other than $\als$ as we shall discuss
later on.
\subsection{Quark masses and nonperturbative corrections}
\nin
 For a typical value of
$R_{\tau}=3.6$ and within the standard SVZ-expansion,
the quark mass and nonperturbative corrections, in units of $10^{-3}$,
are \cite{BNP}:
\bea
\delta^{(2)}&\simeq& -(10 \pm 2) \nnb \\
\delta^{(4)}&\simeq& -(3.3 \pm 0.5) \nnb \\
\delta^{(6)}&\simeq& -(7 \pm 4) \nnb \\
\delta^{(8)}&\simeq& 0.01.
\eea
The one-instanton and instanton-anti-instanton manifest as operators of
dimensions larger or equal than 9. Their respective contributions
 $\delta_I\simeq 10^{-6}$ \cite{NASON} and $\delta_{I-I}
\simeq 10^{-3}$ \cite{BRAUN}
are negligible. However, there
is an internal inconsistency of the approach for large value of $\als$.
A recent
 phenomenological fit \cite{ELASH} has shown
that the instanton effect to $R_\tau$ cannot exceed $0.5 \times 10^{-3}$.
\nin
It is clear from the previous discussions
that these nonperturbative corrections are small compared with the
typical
perturbative corrections
$\delta^{(0)} \simeq 20\%$. This
is due to the fact that the largest quark mass effect due to the
strange quark is suppressed by the sin$^2\theta_c$
factor. Charm and beauty quark masses only contribute via  higher order
$\als^2$
virtual loops, whose effects $\delta^{(2)}_c \simeq 0.4 \times 10^{-3}$
\cite{CHET}
are negligible, while the dimension-four condensate only contributes
through
radiative corrections due to the particular $s$-structure
of $R_\tau$ and the Cauchy theorem.
Higher dimension operators of dimension $D$, including
instantons effects are suppressed by $1/M^D_{\tau}$ and are clearly
suppressed.
Moreover, recent experimental measurements \cite{DUFLOT,CLEO} of the
dimension-four, -six and -eight condensates are consistent with the QCD
spectral sum rules
estimates used to get the theoretical estimate in Eq. (12)
and which indicate the internal consistency of the SVZ-approach.
The sum of the
previous nonperturbative
effects has been also measured by the ALEPH group \cite{ALEPH}:
\beq
\delta_{SVZ}\equiv  \sum^{ }_{ D=4,6,...}\delta^{ (D)} \simeq (3\pm 5)
\times 10^{-3},
\eeq
which confirms the smallness of the nonperturbative contributions
estimated previously.
This tiny
nonperturbative effect, which is of the same size as the $\als^3$
perturbative one,
then permits the extraction of $\als$ with a high accuracy from $R_\tau$.
Comparatively to the other low-energy and deep inelastic
processes used to extract $\als$ \cite{BETHKE,PDG}, tau decays is the
only place
(at present), where the different sources of theoretical contributions
have been
discussed in detail and pushed so far (perturbatively and
nonperturbatively), such that one can have a $solid$ control of different
 theoretical
errors.
\section{THE VALUE OF $\als$}
\nin
Using the different corrections discussed in detail in the previous
works and
summarized in the previous section, and using
the recent PDG94 \cite{PDG} published experimental value
quoted previously,
one obtains from the inclusive mode the value:
\beq
\als(M_\tau)= 0.33 \pm 0.03,
\eeq
as can be deduced from Table 1 \cite{PICH}. The determination
from the recent preliminary data gives:
\bea
\als(M_\tau)&=& 0.355 \pm 0.021~~~~~~~~~ \mbox{ALEPH} \nnb \\
&=& 0.309 \pm 0.024.~~~~~~~~ \mbox{CLEO II},\nnb \\
\eea
with a $preliminary$ average consistent with the PDG94 value. We shall
adopt in the
following discussion the PDG94 value.
\nin
One can also extract the value of $\als$
from the separate axial-vector $R_{\tau,A}$
and vector $R_{\tau,V}$ channels and from the sum of the exclusive modes
$
R_{\tau,excl}\equiv R_{\tau,A}+
R_{\tau,V}+R_{\tau,S}$, where $ R_{\tau,S}$ is the sum of the Cabibbo
suppressed
channel. The compilations of these rates have been done in
\cite{PICH}, while their updated values are given in Table 2 from
\cite{SN94}.
It is interesting to notice that these values are about the same as the
ones from the
latest data \cite{PICHM}. We have used here $R_{\tau,S}=0.145\pm 0.020$
from \cite{SN94}
which agree within the error with the recent data 0.16$\pm 0.02$ of the
direct sum of
the exclusive modes \cite{ALEM}.
We have retained in the average the most accurate error from $R_\tau$.
One should
notice that the relative inaccuracy of $\als$ from the vector and
axial-vector channels
is mainly due to the larger effects of the nonperturbative $D=6$
condensates in these
channels, which are smaller for $R_\tau$.

\begin{table}[hbt]
\begin{center}
\setlength{\tabcolsep}{1.5pc}
\caption{ Values of $\alpha_s$ from different observables
in $\tau$-decays.}
\begin{tabular}{c|c}
\hline
& \\
Observables &  $\alpha_s(M^2_\tau)$ \\
\hline \\
& \\
$R_\tau=3.56 \pm 0.03$ & $0.33 \pm 0.03$ \\
$R_{\tau,V}=1.78 \pm 0.03$& $ 0.35 \pm 0.05$ \\
$R_{\tau,A}=1.67 \pm 0.03$& $ 0.34 \pm 0.05$ \\
$R_{\tau,excl}=3.58\pm 0.05$& $ 0.34 \pm 0.04$ \\
& \\
\hline
\hline
&\\
Average&$0.337 \pm 0.030$ \\
& \\
\hline
\end{tabular}
\end{center}
\end{table}

The previous precision in the determination
of $\als$ has been possible by the very inclusive nature of the
$\tau$-decay process
and by the fact that $M_\tau$ is in compromise energy region where the
process is
low enough in energy for being very sensitive to $\als$ but still high
enough for being less
sensitive to the nonperturbative terms which contribute as high powers
in 1/$M_\tau$,
and which are exceptionnally small due to the
particular $s$-structure of the decay rate and of the Cauchy theorem. In
addition,
complications related to hadronization...are not present here
\footnote{Tau decay is a $lucky$ process
as stated by G. Veneziano.}.
In this respect $\tau$-decay
is an $unique$ process.
 Running the previous average
value of $\als$ until $M_Z$, one obtains:
\beq
\als(M_Z)= 0.121 \pm 0.003.
\eeq
This result indicates that a modest accuracy at the $\tau$-mass
translates into a high-precision
measurement at the $Z^0$-scale as the error shrinks as $\als^2$ when the
scale increases. The
extraordinary agreement with the direct measurement of $\als$ at the
$Z^0$-mass
is a strong indication of the validity of the scale-dependence of the
QCD coupling and of the
asymptotic freedom property of QCD. More speculatively, it can also
indicates that there are
no sizable contributions beyond the standard model (gluino,...) in the
energy region below the
$Z^0$-mass.
\section{THE DIFFERENT SOURCES OF ERRORS TO $\als(M_Z)$}
\nin
Let us now discuss in detail the different theoretical contributions to
the error 0.003 of
$\als(M_Z)$:
\subsection {Running from $M_{\tau}$ to $M_Z$}
\nin
Running the determined value of $\als$ from the $\tau$ to the $Z$-masses
and taking into
account the crossing of heavy flavours \`a la Bernreuther and Wetzel
\cite{BERN},
by setting $\als^f=\als^{f-1}$, when one crosses the threshold of the
$f$-th quark, one
induces an error of 0.0008. We have used the recent determination of the
running $b$ and
$c$ quark masses in \cite{MASS}:
\bea
\overline{m}_b(4.62 ~\mbox{GeV})&=& 4.23^{+0.03}_{-0.04}\pm 0.02~
\mbox{GeV}\nnb \\
\overline{m}_c(1.42~\mbox{GeV}) &=& 1.23^{+0.02}_{-0.04} \pm 0.03~
\mbox{GeV},
\eea
from the bottomium and charmonium sum rules. However,
a more conservative estimate \cite{SN94,BERN,LEDIB}, using quark mass
values with errors as large as $\pm$ 0.3 GeV, leads to an error on
$\als(M_Z)$
of 0.001, indicating that the error $\pm 0.002$ quoted in \cite{ALTA} is
unrealistic.
\subsection{Quark masses and nonperturbative effects}
\nin
The error induced by the nonperturbative effects is dominated by the one
due to the
dimension-6 condensates. With the previous value of $\delta^{(2)}$ and
$\delta_{SVZ}$ estimated in Eq. (12), one induces an error of 0.0008 in
$\als(M_Z)$.
\subsection{RS- and $\mu$-dependence}
\nin
In \cite{LEDI}, Pich and Le Diberder have shown that the perturbative
QCD series is more convergent than the original one in \cite{BNP}, if
one uses the
expansion in terms of the {\it contour coupling} $A^{(n)}$:
\bea
A^{(n)}(a_\zeta)&=& \frac{1}{2i\pi}\oint_{|s|=M^2_\tau}\frac{ds}{s}
\nnb \\
&&\ga 1-2\frac{s}{M^2_\tau}+
2\frac{s^3}{M^6_\tau}-\frac{s^4}{M^8_\tau} \dr a^n_\zeta (s), \nnb \\
\eea
instead of the usual QCD coupling $a_\zeta(s) \equiv \als(\mu)/\pi$,
where $\mu \equiv \zeta M_\tau$, is the introduced subtraction scale.
In terms of
this coupling, the perturbative series becomes:
\bea
\delta^{(0)}_{BNP}&\equiv& \sum_{n=1} K_{n} A^{(n)}(a_\zeta) \nnb \\
&\equiv& \sum_{n=1} \ga K_{n}+g_n\dr a^n_\zeta,
\eea
where $K_n$ are nothing else than the coefficients of $a^n(s)$ of the
${\cal D}$-function given
in Eq. (10),
while $g_n$ (for $SU(3)_f:~g_1=1,~g_2=3.563$ and $g_3$=19.99), which
depends
on the coefficients of the $\beta$-function, are induced by
the running of the QCD coupling. One should notice that $g_n$ is much
larger than $K_n$,
 which explains the large scheme- and scale-dependences of the original
 series expanded
in terms of $a^n$.
\nin
One way to study the Renormalization Scheme (RS) dependence of the
result is to move the
scheme-dependent coefficient $\beta_3$ of the $\beta$-function around
its $\overline{MS}$
value. Using the
$improved$ series, one finds that the determined value of $\als$ is
stable versus the
variation of $\beta_3$ in the range from 0.5 to 2 times its $
\overline{MS}$ value, while
the effect of the variation of $\mu$ is almost negligible for $\mu$
larger than
1.2$M_\tau$. The RS-
and $\mu$-dependences induce respectively an error of 0.0005 and 0.0009
to $\als(M_Z)$
for a given value of $\delta^{(0)}$ of about 0.2.
\subsection{Higher-order perturbative effects}
\nin
In the previous paper \cite{BNP}, the size of the perturbative errors
have been estimated
from a geometric series estimate of the uncalculated $\als^4$
coefficient, which leads to
the estimate of $\pm 130$ of this coefficient. In another
work, this estimate has been
improved to be \cite{LEDI}:
\beq
\delta^{(0)}_4 \simeq (78\pm  25) \left({\alpha_ s \over \pi} \right)^4,
\eeq
where the first term $g_4\equiv 78$ has been induced by the running of
the previous terms
via the Cauchy integral . The error has been estimated by assuming a
geometric
growth of the coefficients of the ${\cal D}$-function ($K_4 \simeq \pm
K_3\ga K_3/K_2\dr$), while it has been multiplied by a factor 2 in
\cite{PICH,SN94} in order to be conservative.
The previous error has been motivated in order to take into account all
unknown higher-order
effects including the ones induced by the summation of the QCD series at
large order
(renormalons, ...).
 With this $conservative$ error estimate,
an error of 0.0014 to $\als(M_Z)$ is then induced.
\nin
However, one could further improve \cite{KATA} the estimate of the
$\als^4$
term of the ${\cal D}$-function by using the PMS \cite {PMS} and
Effective Charge \cite{EC}
schemes approach or by  measuring it directly 	from the dat
 \cite{LEDI94}.
In this way, one obtains respectively the value $K_4 \simeq 28$ and
$29\pm 4 \pm 2$,
which confirm
the previous value $\pm 25$. Therefore,
one can write:
\beq
\delta^{(0)}=\delta^{(0)}_{BNP}+103 \left({\alpha_ s \over \pi} \right)^4
\pm 2\times 94  \left({\alpha_ s \over \pi} \right)^5,
\eeq
where the last coefficient is a bold-guess estimate based again
on the geometric growth of the series, which we have multiplied by a
factor 2 for a conservative estimate. For a typical value of $\als$,
this error
is about 0.4 of the one which we have retained previously, and indicates
that the previous  error
$\pm 50 (\als/\pi)^4 $ is a realistic conservative estimate of the
perturbative error.
\nin
Then, we conclude from our previous discussions that we shall adopt, as
a $conservative$
estimate of the uncertainties related to the truncation of the
perturbative series
including the ones induced by its summation at large order, the value:
\beq
\Delta\delta^{(0)} \simeq \pm 2\times (K_4\simeq 25) \ga\frac{\als}{\pi}
\dr^4.
\eeq
\subsection{The sum of the errors within the SVZ-expansion}
\nin
Adding the previous different sources of errors quadratically,
a total value of 0.002 of the $theoretical$ errors for $\als(M_Z)$
is obtained. The different
errors to the predicted value of $R_\tau$ for various values of $\als$
using the
previous estimates of the errors are given in  Table 1.
\section{SOME SPECULATIVE SOURCES OF ERRORS}
\nin
Now, let me discuss some other possible exotic sources of errors
not included into  the SVZ-expansion, and, at the same time, let me
answer some (unjustified) criticisms raised in the literature. Here, one
should
notice that, in contrast with the previous
$true~quantitative$ estimate of the errors, the discussions are quite
speculative
and very qualitative. We shall argue that our $conservative$ estimate of
the errors
in Eq. (22) contains already the following exotic errors.
\subsection{IR and UV renormalons}
\nin
Renormalons effects are associated to the insertion of n bubbles of
quark loops into
a gluon line exhanged in the ${\cal D}$-function given
in Eq. (9) built from the quark current. It is
well-known that they induce a n! growth into the perturbative series.
This disease can
be (in principle) cured by working with the Borel transform
$\tilde{\cal{D}}$ of the
correlator ${\cal D}(s)$:
\beq
{\cal D}(a\equiv \als/\pi)= \int_{0}^{\infty} db ~\tilde{\cal D}(b)~
exp (-b/a),
\eeq
which possesses an explicit 1/n! suppression factor. However, the life
is not so simple
as $\tilde{\cal D}$ develops
singularities at $b= 2k/(-\beta_1)$ in the real axis, which make the
integral ill-defined.
\nin
The infrared (IR) renormalons which correspond to the singularities
at $k=+2,~+3,...,$ are generated by the low-energy behaviour of the
diagrams,
and can be absorbed into the definitions of the
QCD condensates \cite{AMU}-\cite{GRUN}.
It has been also proved \cite{BENE} that there cannot be a $k=+1$
singularity  and
then, no $1/s$-ambiguity can be generated by the IR renormalons; that is
mainly
related to the absence of the $D=2$ gauge invariant condensates in QCD.
\nin
The  ultraviolet (UV) renormalon singularities corresponding to
$k=-1,~-2,...,$ are generated
by the high-energy behaviour of the virtual momenta and leads to a
Borel-summable series.
After a Borel-sum, they cannot limit the applicability of  perturbation
theory
\cite{AMU,WIL}, though they can
induce an uncertainty in the truncated perturbative series when the
Borel-sum is not done.
Their contributions are dominated by the singularity at $k=-1$, which
are largely
renormalization-scheme dependent.
effects not
The existence of this effect is still controversial, as
according to \cite{AMU,WIL}, this effect should not be present. However,
 if one takes for granted the validity of the
theoretical estimate obtained in the limit of infinite numbers of
flavours
and if one considers the result
from one exchanged gluon-chain with large numbers of fermion
blobs, one would expect \cite{ZAK}\footnote{However, a recent analysis
\cite{VAIN}
has shown that the effects of two chains of gluons can be of the same
order as the
one due to one gluon-chain This feature invalidates the previous result
within a one gluon-chain
approximation}:
\beq
\delta_{UV} \sim A(\mu)\sqrt{\als(\mu)}\ga \frac {\Lambda M_\tau}{\mu^2}
 \dr ^2,
\eeq
where $\mu$ is the subtraction point and $A$ is a renormalization
scheme-dependent coefficient. The result indicates that for $\mu$
larger than $ M_\tau$, the
effects become negligible. Indeed, rigorously, this new term, which is
$\mu$-dependent,
should also be taken into account in the minimization of the
$\mu$-dependence of the QCD
perturbative series discussed previously by \cite{LEDI}. However, it is
incorrect to
say \cite{ALTA} that the result of \cite{LEDI} is false.
Indeed, the effect of this UV term
is only relevant in the region of  $\mu $ smaller than $M_\tau$, where
the stability of the
perturbative series in the change of $\mu$ is not yet reached. In fact,
for $\mu$ larger
than $ M_\tau$, the UV-renormalon effect obtained in this way goes
quickly to 0 like
1/$\mu^4$, so that it
 cannot disturb the stability analysis in $\mu$ performed by \cite{LEDI}.
Moreover, one can also argue \cite{PICHM} that for arbitrary values of
$\mu^2=s$,
the effect of this term vanishes to leading order, like any other
dimension-four terms,
as a consequence of the Cauchy theorem and of the $s$-structure of the
decay rate.
Therefore, it can $never$ introduce a large error in $R_\tau$.
\subsection{$1/M^2_\tau$ terms}
\nin
\cite{ALTA} also
argues that this $\mu$-dependence can disappear when new terms are
added in the evaluation of the UV renormalon effects, in such a way that
the previous
contribution transforms into:
\beq
\delta_{2} \sim c\ga \frac {\Lambda}{ M_\tau} \dr ^2,
\eeq
 where $c$ is an unknown coefficient that can eventually contain a
 log-dependence. However,
this argument is not quite true as the transition from Eq. (24) to Eq.
(25) is just a
reflection of the renormalization scheme dependence of this UV
renormalon effect,
which manifests in Eq. (25) through $\Lambda$.
competition between the
\nin
Alternatively, one could proceed phenomenologically,
by including an $ad~hoc$ $C_2/M^2_\tau$ term in the SVZ-expansion, but
still,
with the caution that {\it this term may not exist at all}.
Then, by fitting
it, independently of the tau decay process, for instance,
from
the  $e^+e^- \rar I=1$ hadrons data on the total cross-section,
as done, for the first time, in \cite{SN92}, one can deduce the quite
inaccurate
optimal estimate:
\beq
|C_2| \simeq (9\pm 4)\times 10^{-3}~\mbox{GeV}^2,
\eeq
from the stability analysis of the
Laplace sum rule\footnote{For those who are not familiar
with the sum rule analysis, the stability
corresponds to the minimum sensitivity of the result with respect to the
variation of the unphysical sum rule variables. The meaning of the
stability
procedure has been
tested in quantum mechanics by applying the sum rule for the harmonic
oscillator eigenvalue
equation \cite{BERT} (see also \cite{SNB} for a review), where both the
exact and approximate
solutions are known.	The smooth plateau is only obtained for the exact
 solution, while
for the approximate series, the optimal solution corresponds to the
minimum of a parabola or to an inflexion point}, while the Finite Energy
Sum Rule gives
a very inaccurate value, which we can translate into the $conservative$
upper bound:
\beq
|C_2| \leq (0.374~\mbox{GeV})^2,
\eeq
 The unusual inaccuracy of the sum rules analysis, is mainly due to the
 sensitivity of the
result on the region around 1.4-2 GeV, where the data are quite bad.
The previous estimate leads to:
\beq
\delta^{(2)} \simeq (18\pm 7)\times 10^{-3}.
\eeq
The error in this result can induce an uncertainty of about 0.002 for
$\als(M_Z)$.
 Better data or an independent  analysis from
other channels than $e^+e^-$ are of course needed for improving the
previous
value of $C_2$. One should notice that the ALEPH group \cite{ALEPH}
has also used the tau-decay
data in order to fix this term, using spectral moments which are much
better that ordinary FESR, but the analysis was not
conclusive.
\nin
 The effect in Eq. (28) is still small compared with the one $proposed$
 in \cite{ALTA}:
\beq
\delta_2 \simeq \frac{1}{(3-5)} \ga \frac{\Lambda \equiv 0.5~\mbox{GeV}}
{M_\tau}\dr^2,
\eeq
which leads to an $overesimated$ uncertainty of 0.005-0.008 to $\als(M_Z)
$. \cite{ALTA}
reinforces his result by comparing it with an alternative derivation of
the error from the
truncation of the perturbative series at the calculated $\als^3$-term,
and by advocating
that the error of the asymptotic series is given by the last known term.
In this way, he
finds:
\beq
\Delta \delta^{(0)}\simeq 10 \ga \frac{\als}{\pi}\dr ^3,
\eeq
leading to an error of about 0.005 to $\als(M_Z)$. There are two amuzing
issues here:
\nin
--It is interesting to notice that \cite{ALTA} recognizes that the errors
 from the $1/M^2_\tau$
in Eq. (29) and from the truncation of the asymptotic series in Eq. (31)
are of the same origin,
so that one should not add them in the total errors.
\nin
--The coefficient of the $\als^3$-term given in Eq. (30) is again an
arbitrary overestimate
compared to the calculated $K_3=6.3711$-coefficient of the
${\cal D}$-function relevant
for the discussion. Indeed, \cite{PICHM} has used a quite similar
argument, based on the
fact that $by~definition$
the UV renormalon effect is smaller than the calculated perturbative
error, and that the
asymptotic behaviour of the perturbative series is reached at order
$\als^3$,
in order to give an estimate of the maximal error due to the truncation
of the perturbative
series. In this way, he obtains:
\beq
  \Delta\delta^{(0)}\leq K_3 a^3 \simeq 0.010.
\eeq
This result is again consistent with the previous one in Eq. (28). Its
consistency with
the estimate of the errors in Eqs (21) and (22) indicates that, not only,
 our $conservative$ error in Eq. (22) is $realistic$, but also, the
 asymptotic series
may already be reached at the $\als^3$ order, though we do not yet have
alternating signs
in the coefficient of the series at this order.
\subsection{Freezing mechanism}
\nin
In \cite{ALT}, it has been suggested that the freezing of the coupling
constant at
low scale can also induce a 1/$M_\tau^2$ contribution, which can be
simply seen by
parametrizing the QCD coupling as:
\beq
\frac{\als}{\pi}{\Big |}_C \simeq \frac{2}{-\beta_1}\frac{1}
{\log{(M^2_\tau+C^2)/
{\Lambda^2}}},
\eeq
where C is an unknown coefficient expected to be smaller than the
hadronic scale $M_\rho$.
Expanding this formula, indeed leads to:
 \beq
 \frac{\als}{\pi}{\Big |}_C \simeq \frac{\als}{\pi}\aga 1-\frac{C^2}
 {M_\tau^2}
\ga \frac{-\beta_1}{2}\dr \frac{\als}{\pi} \adr,
\eeq
which for $C\leq M_\rho$ gives:
\beq
\delta_C \leq 5\times 10^{-3},
\eeq
which is quite small. Again, we notice that the quoted error in
\cite{ALTA} is a factor
two larger than the previous one.
\subsection{Sum of the speculative errors}
\nin
We conclude from our previous discussions that the eventual dominant
contribution from
the $1/M^2_\tau$ terms and the estimate of the perturbative error based
on the truncation
of the series are of the same origin and should not be added in the
total error.
Our $conservative$ estimate of the perturbative error in Eq. (22)
compares quite well with
the $phenomenological$ error
in Eq. (28), which reinforces the previous argument.
\nin
However, though, $grosso$-$modo$, we agree with \cite{ALTA} on the
$philosophical$
interpretation of
the UV renormalon, it appears to us that his handling of the different
numerical errors is
always an overestimate of the $true$ error by about a factor 2-3.
\subsection{Hadronic uncertainties and $e^+e^-$ stability test}
\nin
Let us finally comment on the criticisms raised by \cite{TRUONG} about
the uncertainties
introduced by the hadronic states in connection of the application of
QCD near the real axis.
Our claim is that these effects, which could,
na\"{\i}vely, be important {\it are zero}, thanks to the presence of the
threshold factor $(1-s/M^2_\tau)^2$, which introduces a double zero at
$s=M^2_\tau$.
We have tested \cite{PICH} the validity of our argument by varying the
$\tau$-mass. In so
doing, we express the vector channel decay rate $R_{\tau,V}$ in terms of
the $e^+e^-\rar$
I=1 hadrons total cross-section using CVC. Though these data are at
present less accurate
than the one obtained directly from a measurement of $R_{\tau,V}$, they
are interesting
as they permit to cover a large range of $M_\tau$-values. For $M_\tau$
between 1.4--2 GeV,
the agreement between the theoretical curves and the data on
$R_{\tau,V}$ using the previous
values of $\als$ are excellent. For smaller values of $M_{\tau}$, there
is a large departure
from the theoretical and experimental curves, indicating that we
approach the nonperturbative
regime, where higher-dimensions condensates not retained in the OPE show
up. For larger
values of $M_\tau$, the agreement are still good, but the data are
inaccurate and contradict
with each other which render the analysis in this region unconclusive.
Indeed,
as already emphasized in \cite{PICH,SN94}, the author in Ref.
\cite{TRUONG}
uses $ordinary$ Finite Energy Sum Rules, (i.e. without threshold factors
and so are
more inaccurate than the one gouverning tau decays), in the region
above 1.8 GeV, in order to test the stability of our determination of
$\als$, from which
he found that the $\als$-value can move by a factor more than 2.
However, by examining
carefully his analysis, one can simply realize that
the test of \cite{TRUONG} is more a test of his approach
itself, but is irrelevant for testing the determination of $\als$ from
tau decays.
\section{CONCLUSION}
\nin
We have discussed in detail the different sources of
theoretical errors in the determination of $\als$
from tau decays starting from a typical value
of $R_\tau= 3.56 \pm 0.03$. These errors
have been classified as follows for $ \als (M_Z)$:
\bea
 0.0003&&\mbox{electroweak}\nnb \\
 0.0010&&M_\tau \rar M_Z \nnb \\
 0.0005&&RS~\mbox{-dependence}\nnb \\
 0.0009&&~ \mu~\mbox{-dependence}\nnb \\
 0.0005&&\mbox{quark masses}\nnb \\
 0.0009&&\mbox{SVZ condensates}\nnb \\
 0.0014&&\mbox{truncation of the PT series at} ~\als^4 \nnb \\
&&\mbox{or UV renormalon and }1/M^2_\tau.
\nnb \\
\eea
Adding them in quadrature, we obtain the total $theoretical$ error:
\beq
\Delta \als(M_Z) \simeq  0.0023,
\eeq
while by replacing the perturbative error 0.0014 from
the $\als^4$-term by the maximal value 0.0022 induced by either the
$\als^3$-term
in Eq. (31) or the $phenomenological$ one in Eq. (28), one expects to
have
the phenomenological upper bound:
\beq
\Delta \als(M_Z) \leq  0.0028.
\eeq
We remind that the error for $\als(M_\tau)$ is approximatively ten times
the one
of $\als(M_Z)$.
Our previous detailed analysis has shown that the error of 0.006-0.008
$wished$ by \cite{ALTA} is very
unrealistic as it assumes a huge $1/M^2_\tau$ contribution, where its
numerical derivation is very inaccurate. Moreover, almost all
of the theoretical errors given in \cite{ALTA} have been overestimated by
at a factor 2 to 3.
However, it is amuzing to observe that since his first $attack$ in
\cite{ALT},
his estimate of the error has decreased by a factor of about 2, but
still remains
an overestimate of the $true~quantitative~ error$ estimated in detail
here.
\section*{ACKNOWLEDGEMENTS}
\nin
This talk has been initiated from the collaboration with Eric Braaten
and Antonio Pich
in \cite{BNP}, while during its preparation, I have benefitted from
continuous exchanges with
Toni Pich. Finally, I thank the organizers of this TAU94 workshop for
inviting me to
present this interesting topic.

\end{document}